# Weighting graphene with QCM to monitor interfacial mass changes

Nurbek Kakenov,[1] Osman Balci,[1] Omer Salihoglu,[1]

Seung Hyun Hur,[2] Sinan Balci,[3] and Coskun Kocabas[1,a)]

[1]*Department of Physics, Bilkent University, 06800 Ankara, Turkey*

[2]*School of Chemical Engineering, University of Ulsan, Daehak-ro 93, Nam-gu, Ulsan 680-749, South Korea.*

[3]*Department of Astronautical Engineering, University of Turkish Aeronautical Association, 06790 Ankara, Turkey*

a)*Electronic address: ckocabas@fen.bilkent.edu.tr*

In this Letter, using quartz crystal microbalance (QCM), we experimentally determined the mass density of graphene grown by chemical vapor deposition method. We developed a transfer printing technique to integrate large area single-layer graphene on QCM. By monitoring the resonant frequency of an oscillating quartz crystal loaded with graphene, we were able to measure the mass density of graphene as ~118 ng/cm$^2$, which is significantly larger than the ideal graphene (~76 ng/cm$^2$) mainly due to the presence of wrinkles and organic/inorganic residues on graphene sheets. High sensitivity of quartz crystal resonator allowed us to determine the number of graphene layers in samples. (The technique is very sensitive and able to determine the number of graphene layers in a particular sample.) Besides, we extended our technique to probe interfacial mass variation during adsorption of biomolecules on graphene surface, and plasma-assisted oxidation of graphene. (Besides, we extended our technique to probe interfacial mass variation during adsorption of biomolecules on graphene surface, and plasma-assisted oxidation of graphene.)

2-dimentional (2d) crystals provide a new platform to study physics in reduced dimensions.[1] Graphene is at the center of this new field due to its unique physical properties.[1] Various techniques have been implemented to elucidate the fundamental properties of graphene. For instance, optical spectroscopy has been demonstrated to measure the fine structure constant ($\alpha$) defining visual transparency of graphene.[2] Although the electrical and optical properties of graphene are extensively studied, the mechanical properties of graphene remain much less explored. Graphene yields tunable high mobility charge carriers for electronic and optoelectronic applications; on the other hand, graphene has also other physical properties such as monoatomic thickness together with strong covalent bonds, high stiffness, and low mass density. Due to its low mass density and extraordinary mechanical properties, graphene is an ideal nanomaterial for nano-electromechanical systems (NEMS), e.g., for extremely small mass sensing.[3,4] Understanding basic properties of graphene plays a critical role in improving the performance of nano-electromechanical devices. Accurate measurements of these parameters are essential to model the performance of NEMS devices. For example, the elastic modulus of graphene has been measured to be 1 TPa.[5] A variety of approaches used for synthesis of graphene often produce single or polycrystalline graphene structures thus leading to variations in measured physical properties.[6] For example, mechanical exfoliation yields single crystal graphene whereas chemical vapor deposition (CVD) most frequently yields polycrystalline graphene. Another parameter directly determining performance of NEMS devices is mass density of graphene which is theoretically calculated to be ~76 ng/cm$^2$.[7,8] However, the mass density of CVD grown graphene has not yet been experimentally determined. In modeling of graphene NEMS devices, the value of ~76 ng/cm$^2$ is taken as mass density of graphene[7] but experimentally measured mass density of graphene deviates from its theoretical value of ~76 ng/cm$^2$ since nanoscale wrinkles and (or) molecular adsorbates greatly affect the measured density. Herein we used quartz crystal microbalance (QCM) to experimentally determine mass density of graphene.

Integration of QCM with graphene yields a practical device for measuring mass density and interfacial mass change of graphene. Various surface specific techniques have been implemented to understand these interfacial processes on graphene surface.[9] Mass detection with mechanical resonators is an alternative method with qualities of low cost and extraordinary sensitivity, e.g., recently single nanoparticle weighing has been achieved.[4,10] Graphene[7] and carbon nanotube[11] based nano-electromechanical resonators have been demonstrated for extremely small mass detection. This configuration yields large quality factors of 10,000-100,000.[12,13] Owing to low cost and conceptual simplicity, quartz crystal microbalance provides sensitive means of in situ determination of interfacial mass change.[14] QCM consists of a thin piezoelectric quartz crystal sandwiched between two metal electrodes. Application of an alternating voltage to the electrodes couples electric field to the mechanical oscillations resulting in a vibrational motion of the crystal at the resonance frequency. The quality factor of a QCM can exceed 100000 making QCM an ideal oscillator for sensing extremely small mass variations.[13]

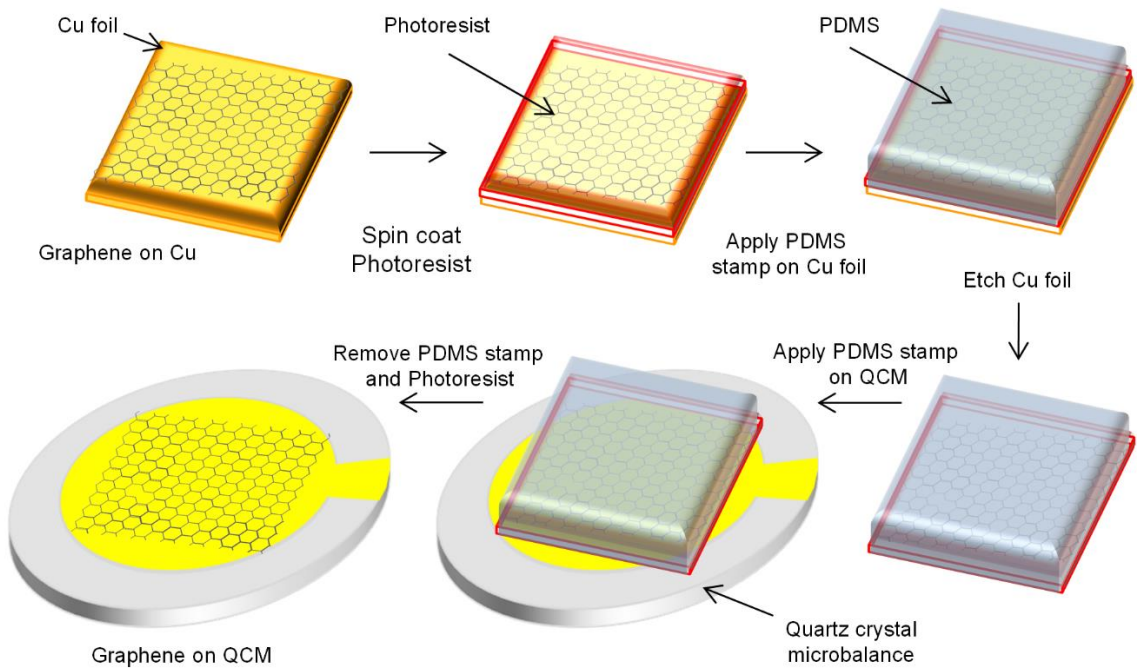

FIG. 1. Transfer printing process of graphene on the front electrode of a QCM.

We integrated QCM with graphene using a transfer-printing process, Figure 1.[15] We used AT-cut, α-quartz crystal with a mechanical response frequency of 5 MHz. The exposed area of electrode is ~1.37 cm$^2$. The area of printed graphene is 0.64±0.05 cm$^2$. Graphene was synthesized by CVD as indicated in our previous works.[16] Graphene coated copper foils were spin coated with a photoresist (PR, AZ5214). A flat elastomeric stamp (PDMS) was placed on the PR layer and the copper foil was etched in the 1 M iron chloride. The stamp was applied to QCM and heated to 100 °C in order to release graphene-PR layer.

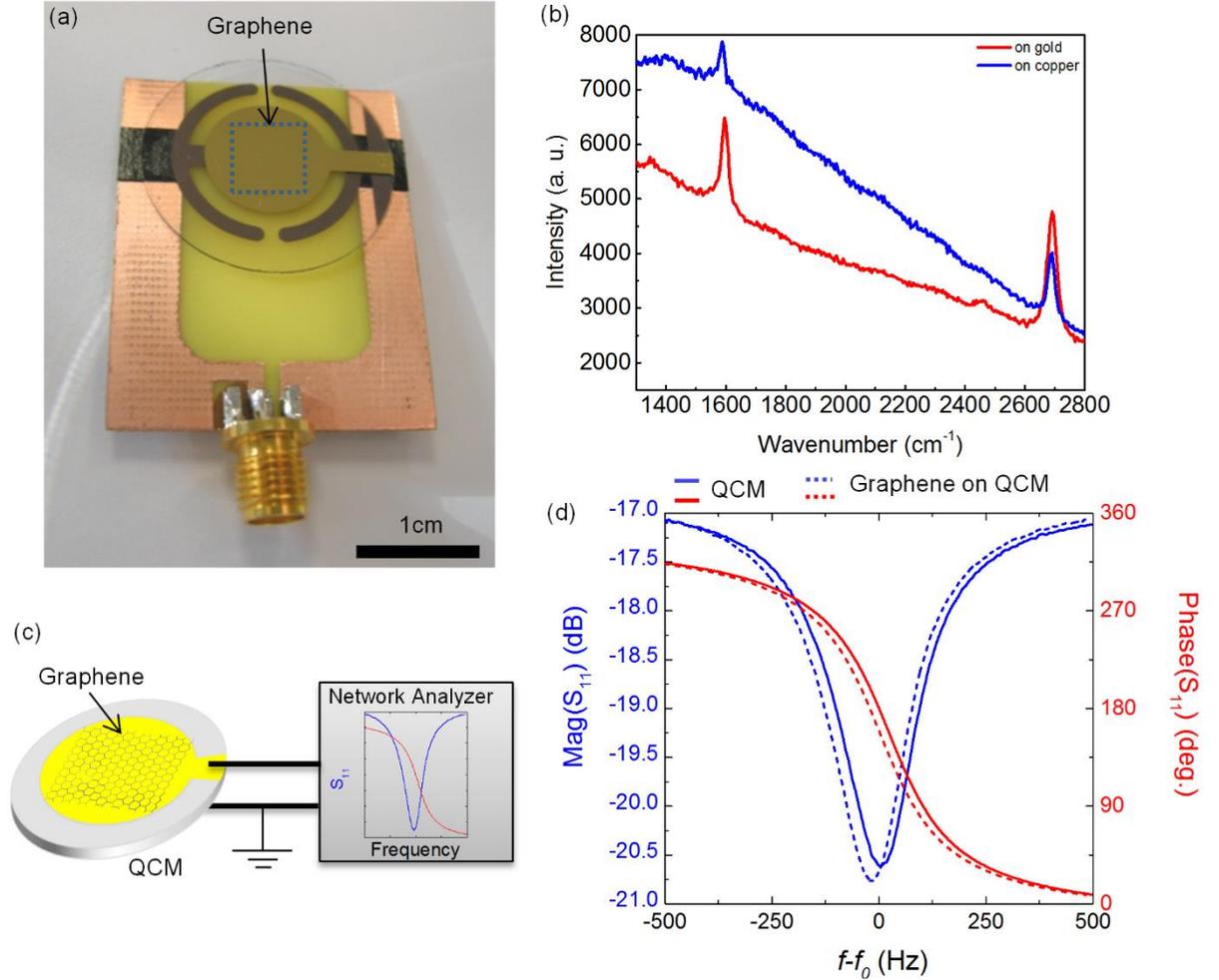

FIG. 2. (a) A photograph of a QCM mounted on a printed circuit board. (b) Raman scattering spectra of graphene on gold and on copper. (c) Experimental setup used for probing resonance characteristic of the QCM. (d) Magnitude and phase of measured scattering parameter $S_{11}$ of port-1, as a function of frequency for blank (solid line) and graphene coated QCM (dot line). The resonance frequency is 5,007,323.4 Hz. The Q-factor of the resonator is ~22600. After coating the surface of QCM with 0.64 $cm^2$ graphene, we observed ~20 Hz shift in the frequency.

Graphene coated QCM was mounted on a printed-circuit-board (PCB), Figure 2(a). We fabricated a PCB and mounted the QCM on PCB. To verify graphene quality, we measured Raman scattering spectra of graphene sheets on copper substrate and on the gold electrode of QCM, Figure 2(b)[17]. The intensity ratio of 2D/G is ~3.4 and the intensity of the defect mode (D) is negligible indicating high quality graphene on QCM. To obtain resonance characteristics of QCM, we measured scattering parameters ($S_{11}$) using a two-port vector network analyzer (VNA, HP 8753D),

Figure 2 (c). S-parameters elucidate the electromechanical properties of QCM and the interaction with the surrounding medium. One port of the VNA is connected to QCM mounted on a crystal holder with a BNC connector. We measured S-parameters for a frequency range of 1 kHz at 5 MHz where the magnitude of $S_{11}$ reaches a minimum value since the total impedance attains its minimum value, Figure 2(d). Next, we transferred a single layer graphene and measured $S_{11}$, Figure 2(d). The area of graphene is 0.64 cm$^2$, which is smaller than the area of the front electrode 1.37 cm$^2$. After graphene transfer, resonance frequency decreases by 20.5 Hz.

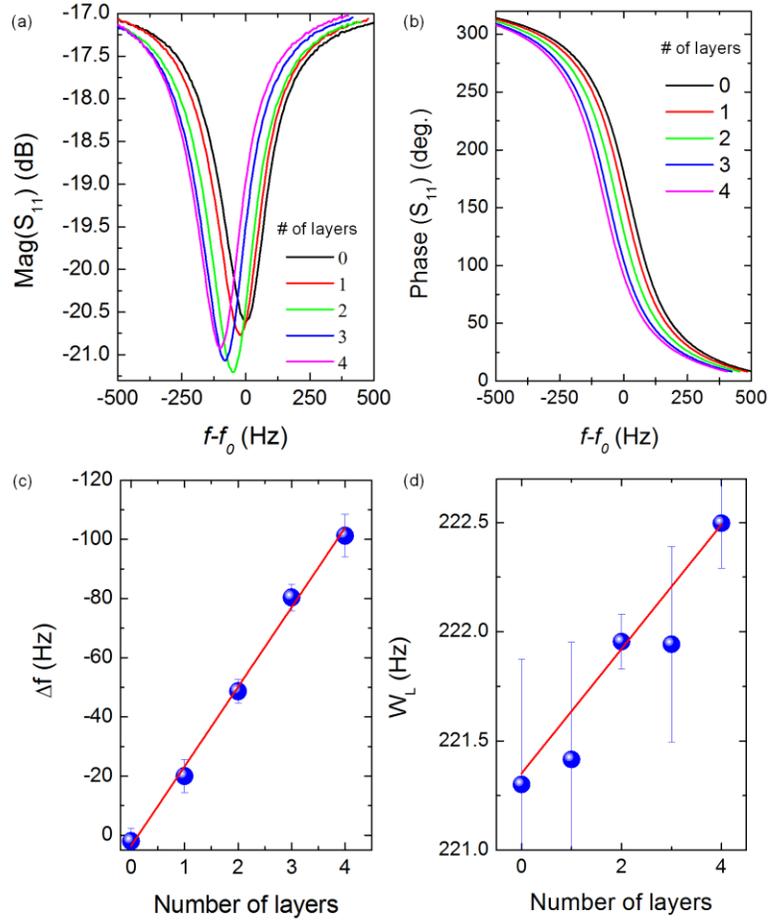

FIG. 3. Magnitude (a) and phase (b) of scattering parameters measured for QCM with multilayer graphene. (c) Measured frequency shift as a function of graphene layer numbers. (d) The bandwidth of the QCM vs. number of graphene layers.

In addition, we integrated QCM with multilayer graphene and measured scattering parameters, Figures 3(a) and 3(b). We found that the resonance frequency linearly increases with increase in number of graphene layers. The scattered plot in Figure 3(c) demonstrates the variation of the resonance frequency with the layer number. The red line shows a curve obtained by linear least squares fitting. The slope of the fitting curve provides the interfacial mass change of 26 ±2 Hz for each graphene layer. Figure 3(d) shows the bandwidth of the QCM as a function of graphene layers.

We then quantified the density of graphene from the measured frequency shifts. The resonance frequency of the crystal is sensitive to variation in interfacial mass as summarized in Sauerbrey's equation.[18]

$$\Delta f = -\frac{2n f_0^2}{\sqrt{\rho_q \mu_q}} \Delta m \qquad (1)$$

where $f_0$ is the resonant frequency of quartz crystal, $n$ is the number of the harmonic at which the crystal is driven. For a 5-MHz, AT cut quartz crystal, the mass density is $\rho_q$=2.648 gcm$^{-3}$, and the shear modulus is $\mu_q$=2.947x10$^{11}$g cm$^{-1}$s$^{-2}$. Here $\Delta m$ defines the change in mass. Sauerbrey's equation, however, provides an average mass density, and in most cases, necessitates calibration. To quantify the frequency change and the mass uptake, we calibrated the QCM by a known amount of mass. A gold film with a thickness of 20 nm was evaporated on QCM (S3). We measured a shift of 4.80 kHz in resonant frequency. The density of the evaporated gold thin film is 16 g/cm$^3$ corresponding to the mass uptake of 20.5 µg.[19] For an exposed area of 0.64 cm$^2$, the calibration factor is 4.27 ng/Hz. For each graphene layer, we obtained 26 Hz frequency shift corresponding to a mass uptake of 111 ng. It should be noted here that this value includes both the mass of graphene and organic/inorganic residues. The extent of the residual contamination on graphene has been characterized by employing variety of techniques.[20] Besides, we performed additional experiments to determine amount of chemical residues on the QCM. To elucidate the mass of residues, we performed a similar transfer process without growing graphene on copper foils. The observed frequency shift for each transfer process is 8.3 Hz corresponding to a mass uptake of 35 ng. After subtracting the residue mass from the measured mass and normalizing by the area, we estimated the mass density of graphene as ~118 ng/cm$^2$. This value is significantly larger than the theoretical value of ~76 ng/cm$^2$ calculated from the mass density of bulk graphite.[7] The larger mass density of CVD graphene is likely due to the wrinkles formed on graphene.[21] Previously, high densities of wrinkles were observed in large-scale growth of graphene on metallic substrates.[21] Figure 4(a) and 4(b) show the scanning electron and atomic force microscope images of a wrinkled graphene flake on copper foils. The period of wrinkles is around 200 nm. These wrinkles are formed due to thermal contraction of the substrate during the cooling process. These high density wrinkles increase the surface area and mass density of CVD graphene. Besides, nanoscale holes, cracks, organic/inorganic residues alter the measured mass density of graphene. There are also other sources of errors originating from the calibration procedure, e.g., the thickness and mass density of gold film used for the calibration and uncertainties in the measured area of graphene.

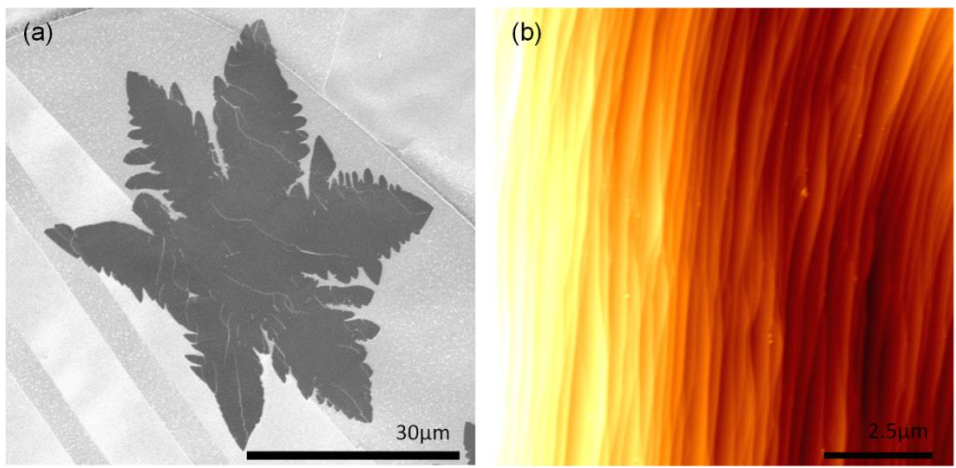

FIG. 4. (a) Scanning electron microscope and (b) atomic force microscope images of wrinkled graphene flakes on copper foils. Obviously, nanometer-sized wrinkles are formed on graphene.

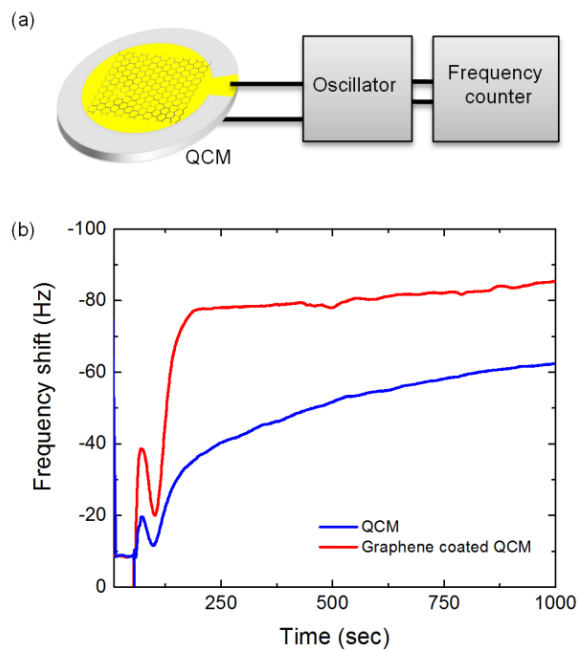

FIG. 5. (a) Schematics of the experimental setup used to probe time trace of the resonance frequency of the QCM. (b) Overlaid time trace of resonance frequency indicating the binding kinetics of BSA (100 nM) on bare and graphene coated QCM surfaces.

We now describe application of graphene modified QCM mass sensor in biomolecule detection, Figure 5(a). Recently, single nanoparticle mass and position have been simultaneously measured using a nanomechanical resonator.[10] By integrating the QCM sensor (Stanford Research, QCM200) with a flow chamber, we monitored the binding dynamics of proteins on graphene.[22] Quartz crystal is placed in a feedback loop of an oscillator. The time trace of the resonance frequency was monitored by means of a frequency counter. The kinetic parameters of the nonspecific binding on a surface can be extracted from the time trace of the resonance frequency. We used 100 nM of Bovine serum albumin (BSA) protein to study nonspecific binding on bare and graphene coated gold electrodes. The time constants, obtained by fitting the time trace data, are 390 sec and 30 sec, for bare and graphene modified QCMs, respectively. From the measured the time constants, we estimated the association constants of BSA on bare and graphene coated gold surfaces as 0.26 $M^{-1}sec^{-1}$ and 3.3 $M^{-1}sec^{-1}$, respectively.

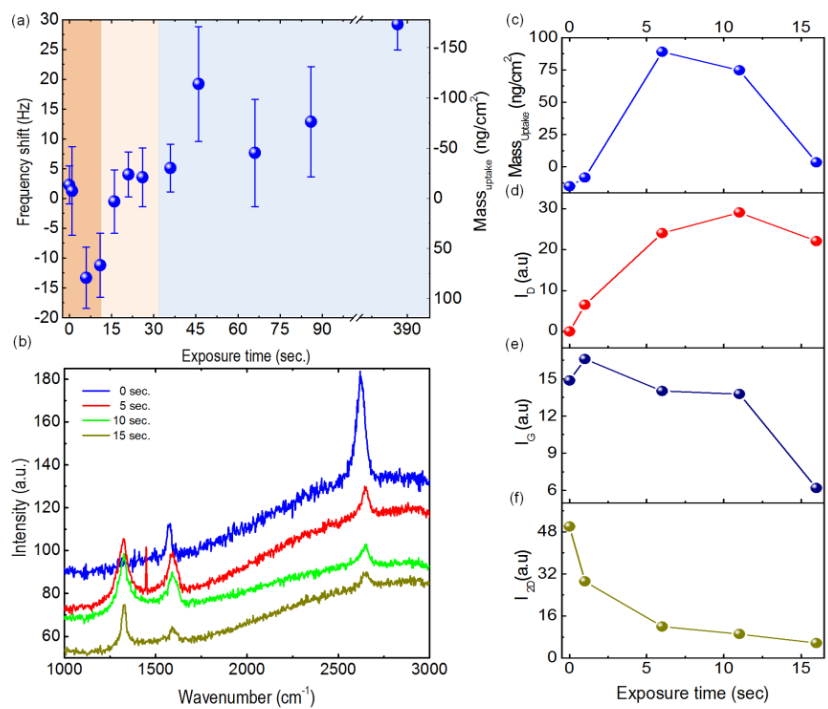

FIG. 6. Probing oxidation of graphene by mass detection. (a) Variation of the resonance frequency and associated mass uptake of graphene exposed to oxygen plasma. (b) Raman spectra of graphene with various exposure times. (c-f) Mass uptake of graphene and correlated Raman intensity for D, G, and 2D bands.

In addition, sensitive mass detection enabled by QCM can be used for monitoring chemical reactions on graphene. Therefore, we traced oxidation of graphene by monitoring the interfacial mass change and correlating the mass change with Raman spectroscopy. QCM with graphene was exposed to oxygen plasma, Figure 6. It is well known that graphene undergoes chemical oxidation when exposed to plasma.[23] We directly correlate the mass uptake with the structure of graphene using Raman spectrum of graphene after each oxygen plasma exposure, Figure 6(b). In the diagram in Figure 6(a), we observe three different regimes. In 10 sec. exposure, interfacial mass increases by 93 ng/cm$^2$, Figure 6(c). This mass uptake is most likely due to the oxidation of graphene and adsorption of other molecules from the plasma. The intensity increase in the D-band (1950 cm$^{-1}$) confirms the formation of sp bonds, associated with lattice distortions. The decrease in intensity of 2D-band (2700 cm$^{-1}$) indicates deformation of graphene. After 15 sec. exposure, a drastic decrease was observed in the interfacial mass, which is most likely due to $CO_2$ release that is expected for the early stage of graphene oxidation. When we further increased the exposure time, we observed complete removal of graphene. This step could include physical removing of small graphene/graphene oxide flakes due to bombarding of surface by energetic oxygen molecules. Correlating the interfacial mass change with Raman spectra could provide more useful information for understanding interfacial reactions on graphene, Figures 6(e-f).

In conclusion, owing to its low cost and ease of use, QCM sensor is a practical way to study the interfacial processes on graphene. By integrating QCM with graphene, we were able to experimentally determine mass density of graphene. The discrepancy between the theoretical and experimental values is mainly due to (i) organic residues, (ii) wrinkles, and (iii) water molecules on graphene. In addition, we traced the adsorption of biomolecules on graphene modified QCM in real-time. The hydrophobicity of graphene could be the main dominant effect that enhances the associativity of proteins by more than 12 times. Besides, chemical reactions on the oxygen plasma treated graphene has been monitored. Indeed, this method can be further developed to investigate interfacial mechanisms on graphene.

This work is supported by the Scientific and Technological Research Council of Turkey (TUBITAK) grant no. 113F278. C.K. also acknowledges the support from the European Research Council (ERC) Consolidator Grant ERC – 682723 SmartGraphene.